\documentclass[aps,prb,twocolumn,superscriptaddress,groupedaddress,a4paper]{revtex4}  
\usepackage{graphicx}
\usepackage{dcolumn}
\usepackage{bm}
\usepackage{amssymb}
\usepackage{amsmath}
\usepackage{subfigure}
\usepackage{color}
\usepackage[paperwidth=210mm,paperheight=297mm,centering,hmargin=2cm,vmargin=2cm]{geometry}

\begin{document}

\widetext
\title{Impact of alloy disorder on the band structure of compressively strained GaBi$_x$As$_{1-x}$}


\author{Muhammad Usman}
\email{usman@alumni.purdue.edu}
\affiliation{Tyndall National Institute, Lee Maltings, Dyke Parade, Cork, Ireland}

\author{Christopher A. Broderick}
\affiliation{Tyndall National Institute, Lee Maltings, Dyke Parade, Cork, Ireland}
\affiliation{Department of Physics, University College Cork, Cork, Ireland}

\author{Zahida Batool}
\affiliation{Advanced Technology Institute and Department of Physics, University of Surrey, Guildford, Surrey, GU2 7XH, United Kingdom}

\author{Konstanze Hild}
\affiliation{Advanced Technology Institute and Department of Physics, University of Surrey, Guildford, Surrey, GU2 7XH, United Kingdom}

\author{Thomas J. C. Hosea}
\affiliation{Advanced Technology Institute and Department of Physics, University of Surrey, Guildford, Surrey, GU2 7XH, United Kingdom}
\affiliation{Ibnu Sina Institute for Fundamental Science Studies, Universiti Teknologi Malaysia, Johor Bahru,
Johor 81310, Malaysia}

\author{Stephen J. Sweeney}
\affiliation{Advanced Technology Institute and Department of Physics, University of Surrey, Guildford, Surrey, GU2 7XH, United Kingdom}

\author{Eoin P. O'Reilly}
\affiliation{Tyndall National Institute, Lee Maltings, Dyke Parade, Cork, Ireland}
\affiliation{Department of Physics, University College Cork, Cork, Ireland}

\vskip 0.25cm



\begin{abstract}
The incorporation of bismuth (Bi) in GaAs results in a large reduction of the band gap energy (E$_g$) accompanied with a large increase in the spin-orbit splitting energy ($\bigtriangleup_{SO}$), leading to the condition that  $\bigtriangleup_{SO} > E_g$ which is anticipated to reduce so-called CHSH Auger recombination losses whereby the energy and momentum of a recombining electron-hole pair is given to a second hole which is excited into the spin-orbit band. We theoretically investigate the electronic structure of experimentally grown GaBi$_x$As$_{1-x}$ samples on (100) GaAs substrates by directly comparing our data with room temperature photo-modulated reflectance (PR) measurements. Our atomistic theoretical calculations, in agreement with the PR measurements, confirm that E$_g$ is equal to $\bigtriangleup_{SO}$ for $\textit{x} \approx$ 9$\%$. We then theoretically probe the inhomogeneous broadening of the interband transition energies as a function of the alloy disorder. The broadening associated with spin-split-off transitions arises from conventional alloy effects, while the behaviour of the heavy-hole transitions can be well described using a valence band-anticrossing model. We show that for the samples containing 8.5\% and 10.4\% Bi the difficulty in identifying a clear light-hole-related transition energy from the measured PR data is due to the significant broadening of the host matrix light-hole states as a result of the presence of a large number of Bi resonant states in the same energy range and disorder in the alloy. We further provide quantitative estimates of the impact of supercell size and the assumed random distribution of Bi atoms on the interband transition energies in GaBi$_{x}$As$_{1-x}$. Our calculations support a type-I band alignment at the GaBi$_x$As$_{1-x}$/GaAs interface, consistent with recent experimental findings. 
\end{abstract}



\maketitle


\section{Introduction}
\label{sec:introduction}

There is increasing interest in the highly mismatched semiconductor alloy, GaBi$_x$As$_{1-x}$, both from a fundamental perspective \cite{Deng_PRB_2010, Usman_PRB_2011, Ludewig_JCG_2012, Bertulis_APL_2006} and also because of its potential device applications. \cite{Sweeney_ICTON_2011, Jin_JAP_2013, Broderick_SST_2012, Mazur_Nano_2011} When a small fraction of As is replaced by Bi in GaAs, the energy gap $E_g$ decreases rapidly, by $\approx 90$ meV when $1\%$ of As is replaced by Bi. In addition, photo-reflectance measurements show that the energy separation, $\Delta_{SO}$, between the spin-split-off valence band and the valence band edge also increases rapidly with Bi composition. \cite{Batool_JAP_2012} This then leads to the situation in GaBi$_x$As$_{1-x}$, where the spin-orbit-splitting energy can exceed the energy gap $\Delta_{SO} \geq E_g$ for $x \gtrsim 9\%$. This is of significant potential benefit for telecomm lasers, because it could enable the suppression of the Auger recombination losses which dominate the threshold characteristics of GaInAsP and AlGaInAs lasers.  \cite{Sweeney_ICTON_2011, Sweeney_patent_2010, Broderick_SST_2012}

The strong band gap bowing in GaBi$_{x}$As$_{1-x}$ is similar to that observed in GaN$_x$As$_{1-x}$, where $E_g$ decreases by as much as 150 meV when $1\%$ of As is replaced by N. The band gap reduction in GaN$_x$As$_{1-x}$ has been well explained using a band-anticrossing (BAC) model. \cite{Shan_PRL_1999} It is well established that replacing a single As atom by N introduces a resonant defect level above the conduction band edge (CBE) in GaAs. \cite{Wolford_Proc_ICPS_1984, Liu_APL_1990} This occurs because N is considerably more electronegative and is also $40\%$ smaller than As. The interaction between the resonant N states and the CBE of the host GaAs matrix then accounts for the observed rapid reduction in $E_g$. It was proposed \cite{Alberi_PRB_2007}, and later confirmed theoretically \cite{Usman_PRB_2011, Broderick_PSSb_2012} that replacing As by Bi can introduce a similar BAC effect. Like N and As, Bi is also a group-V element, which lies below Sb in the periodic table. Bi is much larger than As, and is also less electronegative. It should therefore be expected that any Bi-related resonant defect levels should lie in the valence band (VB) and that, if an anticrossing interaction occurs, it will occur between the Bi-related defect levels and the valence band edge (VBE) of the GaAs matrix. Recent $sp^3s^*$ tight-binding calculations which we have performed confirm the existence of such a band-anticrossing interaction, with the calculated variation of $E_g$ and of $\Delta_{SO}$ in excellent agreement with experiment. \cite{Usman_PRB_2011}

Although the BAC model provides an accurate description of some of the key features in the band structure of highly mismatched alloys, it includes several critical simplifying assumptions. It assumes for instance in GaBi$_x$As$_{1-x}$ that each Bi atom introduces resonant defect levels with constant energy $E_{Bi}$, and then treats the effect on the band structure in terms of the interaction between these states and the host matrix valence band. The BAC model therefore effectively assumes that all Bi atoms are in an identical environment. Although this may be approximately true for very low Bi compositions, where most Bi atoms are widely separated from each other, this assumption can be expected to break down with increasing Bi composition. Firstly, there should be an increasing number of Bi pairs and clusters formed with increasing Bi composition, where a Bi pair consists of a Ga atom that has two Bi neighbours, with clusters then containing larger numbers of contiguous Bi atoms. These pairs and clusters introduce defect levels which lie above the isolated Bi state energy, $E_{Bi}$ and which also interact with the valence band states \cite{Usman_PRB_2011}. In addition, isolated Bi atoms also see an increasingly disordered local environment, leading to an inhomogeneous broadening of the associated resonant defect level energies. We investigate here the effects of these two different types of disorder on the valence band structure of GaBi$_x$As$_{1-x}$ grown pseudomorphically on GaAs, using the $sp^3s^*$ tight-binding model introduced in Ref. 2. 

This paper presents a combined theoretical and experimental analysis of the effects of disorder on the electronic structure of GaBi$_x$As$_{1-x}$ epilayers grown under compressive strain on (100) GaAs substrates. Such layers are of interest for a number of reasons. Firstly, many applications such as semiconductor lasers and related devices require the growth of such layers. Secondly, the growth of these strained structures has an interesting impact on the band structure, splitting the degeneracy of the heavy-hole and light-hole states at the valence band maximum, with the splitting between the two sets of states initially increasing for instance at a rate of about $\approx$ 75 meV per \% lattice mismatch for a compressively strained InGaAs layer grown pseudomorphically on GaAs. We note that the compressive strain in a GaBi$_x$As$_{1-x}$ pseudomorphic layer should also split the four-fold degeneracy of the Bi resonant states, with the two states with heavy-hole symmetry shifting upwards in energy, while the light-hole states shift downwards. With increasing strain, the GaAs host matrix light-hole states should start to pass through the Bi heavy-hole-like resonant states. The intrinsic disorder in GaBiAs then lead to mixing between these two types of state.  

The GaBi$_x$As$_{1-x}$ samples studied were grown by molecular beam epitaxy with Bi concentrations of $x$ = 2.3\%, 4.5\%, 8.5\%, and 10.4\%. \cite{Batool_JAP_2012, Lu_APL_2009, Lu_APL_2008} The samples are fully strained (pseudomorphic) to the GaAs substrate, as confirmed by x-ray diffraction data indicating the GaAs in-plane lattice constant for the GaBi$_x$As$_{1-x}$ layers. \cite{Lu_APL_2009, Lu_APL_2008, Tixier_APL_2005} The theoretical analysis was undertaken using the $sp^3s^*$ tight-binding Hamiltonian which we have developed  for GaBiAs \cite{Usman_PRB_2011} to investigate the electronic structure of large randomly disordered supercells. The electronic structure is studied experimentally by photo-modulated reflectance (PR) spectroscopy \cite{Batool_JAP_2012}, which is considered to be an excellent technique due to its sensitivity to critical point transitions in the band structure. \cite{Pollak_book} Further details of the experimental procedure are described in Batool \textit{et al}. \cite{Batool_JAP_2012}, while the sample details can be found in Lu \textit{et al}. \cite{Lu_APL_2008, Lu_APL_2009}

Three sets of features are found in the PR spectra, associated with transitions between the conduction band minimum at $\Gamma$ and valence states which respectively include host matrix heavy-hole (HH), light-hole (LH) and spin-split-off (SO) character. Our calculations provide a detailed understanding of the broadening observed in the PR measurements for each of these features due to the presence not just of isolated Bi atoms but also of pair and cluster states. We find different factors affecting each of the three transitions, with the calculated data being in generally good agreement with the measured spectra.  In order to test the effects of disorder, we undertake a set of supercell calculations for each of the 4 samples investigated, where the supercell size is varied from 1000 to 8000 atoms, with several calculations with different random distributions of Bi atoms being undertaken for each supercell size. We find for each supercell size that the calculated value of the energy gap separating the lowest conduction and highest (HH-like) valence states varies for a given Bi composition $x$ depending on the actual distribution of  Bi pair and cluster states in the supercell considered, consistent with the inhomogeneous broadening of the lowest energy transition in the PR spectra. The calculated transition energy to the SO band is largely independent of the supercell size and configuration, but we find that the GaAs host matrix SO state character is spread over a number of supercell states, contributing also to an inhomogeneous broadening of the transition energy, which increases with increasing Bi composition $x$. Finally we consider transitions to states which include GaAs LH zone-centre character. Because these states become degenerate with a broad distribution of Bi-related resonant states in a strained layer, we find that transitions to the LH states become very strongly perturbed with increasing Bi composition, with the GaAs LH $\Gamma$ character distribution varying both with supercell size and with the particular random distribution of Bi compositions. At higher Bi compositions, no single feature can be associated with the GaAs LH states, consistent with the difficulty in fully fitting the LH part of the spectrum at higher Bi compositions.

The remainder of the paper is organised as follows. We give a brief overview of the theoretical and experimental methodology in section \ref{sec:methodology}. The main results are then presented in section \ref{sec:results_discussion}. We begin by presenting a general comparison between the experimental and theoretical results in section \ref{sec:theory_exp}. This is followed in section \ref{sec:supercell_size} by a more detailed analysis of the effect of supercell size and of assumed Bi atom distribution on the calculated spectra. We consider supercell sizes of 1000, 4096, and 8000 atoms for each of the four grown samples. We find for a fixed Bi composition that a larger supercell size results in an enhanced inhomogeneous broadening in particular of the LH character. The results clearly highlight the importance of performing large supercell calculations to model realistic dilute alloy environments. In section \ref{sec:random_bi}, we quantify the impact of the random distribution of Bi atoms on the broadening of transition energies by performing 8000 atom supercell calculations with four different random distributions of Bi atoms for each of the four compositions considered. Our results indicate a significant broadening of the transition energies due to the random distribution of atoms, showing that the local band edge energies in a GaBiAs alloy are very sensitive to the Bi distribution in the given region. In section \ref{sec:summary_fits}, we summarise the key physical insights gained by comparison of our atomistic simulations with the measured PR spectra. In section \ref{sec:band_alignment}, we consider the band edge alignment in compressively strained GaBi$_x$As$_{1-x}$/GaAs samples. The band alignment at the GaBi$_x$As$_{1-x}$/GaAs interface remains uncertain in the literature with reports suggesting that the conduction band offset is of type I, \cite{Usman_PRB_2011, Nadir_APL_2012, Pettinari_APL_2008} or type II, \cite{Alberi_PRB_2007} or nearly flat. \cite{Tominaga_ISLC_2010} Our calculations support a type-I band alignment for all the samples considered. The overall conclusions of our work are summarised in section \ref{sec:Conclusions}.

\section{Methodology}
\label{sec:methodology}

The atomistic calculations are performed using a tight-binding (TB) model, where the alloy atom positions are determined using a valence force field (VFF) strain energy minimization scheme and the electronic structure is then calculated using an  $sp^3s^*$ TB Hamiltonian, including bond-length and bond-angle dependent interaction parameters. The VFF and TB models are implemented in \underline{N}ano\underline{E}lectronic \underline{MO}deling Simulator (NEMO 3-D). \cite{Klimeck_1, Klimeck_2} The values of the tight-binding parameters for GaAs and GaBi, as well as the values of the force constants ($\alpha$ and $\beta$) for the VFF model are reported in Ref. 2. The VFF parameters for GaBi were obtained by fitting to first principle calculations of GaBi elastic constants. \cite{Martin_PRB_1970} We further verified \cite{Usman_PRB_2011} that the relaxed Ga-Bi bond lengths in GaBiAs calculated using the VFF model were in good agreement with the results of the x-ray absorption spectroscopy measurements, \cite{Ciatto_PRB_2008} thereby confirming the validity of the VFF mdoel employed. The TB and VFF models have previously been applied to study free-standing GaBi$_x$As$_{1-x}$ and GaBi$_x$P$_{1-x}$ supercells containing up to 8000 atoms. These calculations showed that Bi-related resonant states interact with GaAs host matrix valence band states in GaBi$_x$As$_{1-x}$ via a composition dependent band anti-crossing (BAC) interaction, which we calculate to shift the highest valence band edge upwards in energy by $\approx$ 55 meV per \% of Bi replacing As for low Bi compositions. This BAC interaction is analogous to the BAC interaction between nitrogen (N) defect states and GaAs conduction band states in GaN$_{x}$As$_{1-x}$ alloys. \cite{Reilly_SST_2009} The calculated compositional dependence of the band gap and spin-orbit-splitting energies demonstrated excellent agreement with the available experimental data sets including Bi samples with compositions up to 10\%. The experimentally observed large bowing of the band gap energy was explained in terms of two factors: (i) a linear decrease of $\approx$ 30 meV per $\%$ Bi in the lowest conduction band edge energy (E$_{c1}$) which occurs due to a conventional alloy shift and (ii) a large non-linear increase in the highest valence band edge energy (E$_{v1}$) of $\approx$ 30-55 meV per $\%$ Bi due to the band anti-crossing interaction. The energy of the spin-split-off band (SO) was calculated to shift downwards in energy slightly, exhibiting a linear decrease of $\approx$ 5 meV per $\%$ Bi. It should be noted that no evidence was found for any BAC interaction involving the SO band, as earlier assumed by Alberi \textit{et al.} \cite{Alberi_PRB_2007}

In this paper, we apply our TB model to investigate compressively strained GaBi$_x$As$_{1-x}$ samples, grown epitaxially on a (100) GaAs substrate. In the VFF relaxation scheme, we keep the in-plane lattice constant for all supercells considered fixed to that of the GaAs lattice constant, while allowing the supercell to relax freely along the growth direction. We use periodic boundary conditions in all three spatial directions for the calculations of electronic structure. For each calculation undertaken, the Bi atoms are placed on As sites, selected randomly in the supercells.

In order to compare the theoretical calculations with the experimental data, we calculate the distribution of the host matrix $\Gamma$-character, $G_{\Gamma}(E)$, across the alloy valence band states. The calculation of the $\Gamma$-character is carried out by projection of the six highest host matrix zone centre valence band states (two HH, two LH and two SO) onto the full spectrum of the energy states in the alloy (GaBi$_x$As$_{1-x}$) supercell under consideration. In what follows, we use the subscripts $l,0$ and $k,1$ to index the unperturbed host and Bi-containing alloy states, respectively. The $G_{\Gamma}(E)$ spectrum is then calculated by projecting the unperturbed Ga$_{M}$As$_{M}$ supercell states at $\Gamma$, $\vert \psi_{l,0} \rangle$, onto the spectrum of the Ga$_{M}$Bi$_{L}$As$_{M-L}$ alloy supercell, $\left\lbrace E_{k}, \vert \psi_{k,1} \rangle \right\rbrace$:

\begin{eqnarray}
	G_{\Gamma} \left( E \right) &=& \sum_{k} \sum_{l=1}^{g(E_{l})} f_{\Gamma,kl} \; \delta \left( E_{k} - E \right) \label{eq:FGC_spectrum} \\
	f_{\Gamma,kl} &=& \vert \langle \psi_{k,1} \vert \psi_{l,0} \rangle \vert ^{2} \label{eq:FGC_definition}
\end{eqnarray}

\noindent
Here $\delta \left( E_{k} - E  \right)$ is the Dirac delta function centred on $E_{k}$, $g(E_{l})$ is the degeneracy of the host matrix band having energy $E_{l}$ at the $\Gamma$ point in the Brillouin zone and $\sum_{l}f_{\Gamma,kl}$ is the fractional $\Gamma$ character of the alloy state $\vert \psi_{k,1} \rangle$. Since the PR measurements are performed at room temperature (300 K) and the TB calculations are parametrised to the low temperature band structure, we shift the energy scale of our theoretically calculated $\Gamma$-character plots at 0 K by 0.098 eV to convert them to room temperature values.

The analysis of $G_{\Gamma}(E)$ plots has provided a very useful approach to understand the character and evolution of alloy band edge states. \cite{Usman_PRB_2011, Reilly_SST_2009} Analysis of such plots confirms the presence of the band anti-crossing interaction in ordered supercell calculations of dilute bismide alloys. \cite{Broderick_12band_2012}

Turning to the experimental data, the measured PR spectral line shapes were least-squares fitted using Aspnes third derivative functionals of the form  \cite{Aspnes_SurfSci_1973}:

\begin{equation}
  \label{eq:PR_fit}
  \frac{ \bigtriangleup R }{ R } = Re [ C e^{i \theta} \left( E - E_{g} + i \gamma \right)^{-n} ]
\end{equation}

\noindent
where C and $\theta$ are amplitude and phase variables, E the energy of the probe beam, E$_g$ the critical point transition energy, and $\gamma$ a broadening parameter. We investigated the effect of choosing several of the suggested values for the line shape exponent factor $n$, \cite{Aspnes_SurfSci_1973} but found that this had little influence on the fitted transition energies. Here, we present the results for fits using $n = 3$. In the next section \ref{sec:theory_exp}, we use Eq. \ref{eq:PR_fit} to fit the measured PR data and identify the interband transition energies.

\begin{table*}
\vspace{0ex}
\caption{\label{tab:table1} The values of the broadening parameters $\gamma$ used in Eq.~\ref{eq:PR_fit} to fit the PR spectra plotted in Figs. \ref{fig:HH_LH_PR} and \ref{fig:SO_PR}.}
\small{\begin{tabular}
{@{\hspace{1.0ex}}l@{\hspace{0.1cm}}c@{\hspace{1.0cm}}c@{\hspace{1.0cm}}c}
\\
\hline
\hline
Bi composition    & $\gamma_{HH}$  & $\gamma_{LH}$  & $\gamma_{SO}$   \\
(\%)			  & (meV)  & (meV)  & (meV)           \\
\hline
2.3               & 82.1 $\pm$ 5.3 & 51.9 $\pm$ 3.5 & 30.9 $\pm$ 0.3 \\
4.5     		  & 31.3 $\pm$ 1.2 & 31.0 $\pm$ 2.0 & 22.2 $\pm$ 1.3 \\
8.5               & 31.4 $\pm$ 0.5 & 82.1 $\pm$ 7.0 & 31.9 $\pm$ 2.2 \\
10.4              & 31.4 $\pm$ 1.0 & 31.4 $\pm$ 1.0$^*$ & 38.4 $\pm$ 3.2 \\
\hline
\hline
*Here $\gamma_{LH}$ was constrained to equal $\gamma_{HH}$ in the \\ fit for 10.4\% Bi, due to an extra feature in the \\ LH energy range which is difficult to model. 
\end{tabular}}
\end{table*}

\section{Results and Discussions}
\label{sec:results_discussion}

\subsection{Comparison of Theory and Experiment}
\label{sec:theory_exp}

\begin{figure}
\centering
\includegraphics[width=0.5\textwidth]{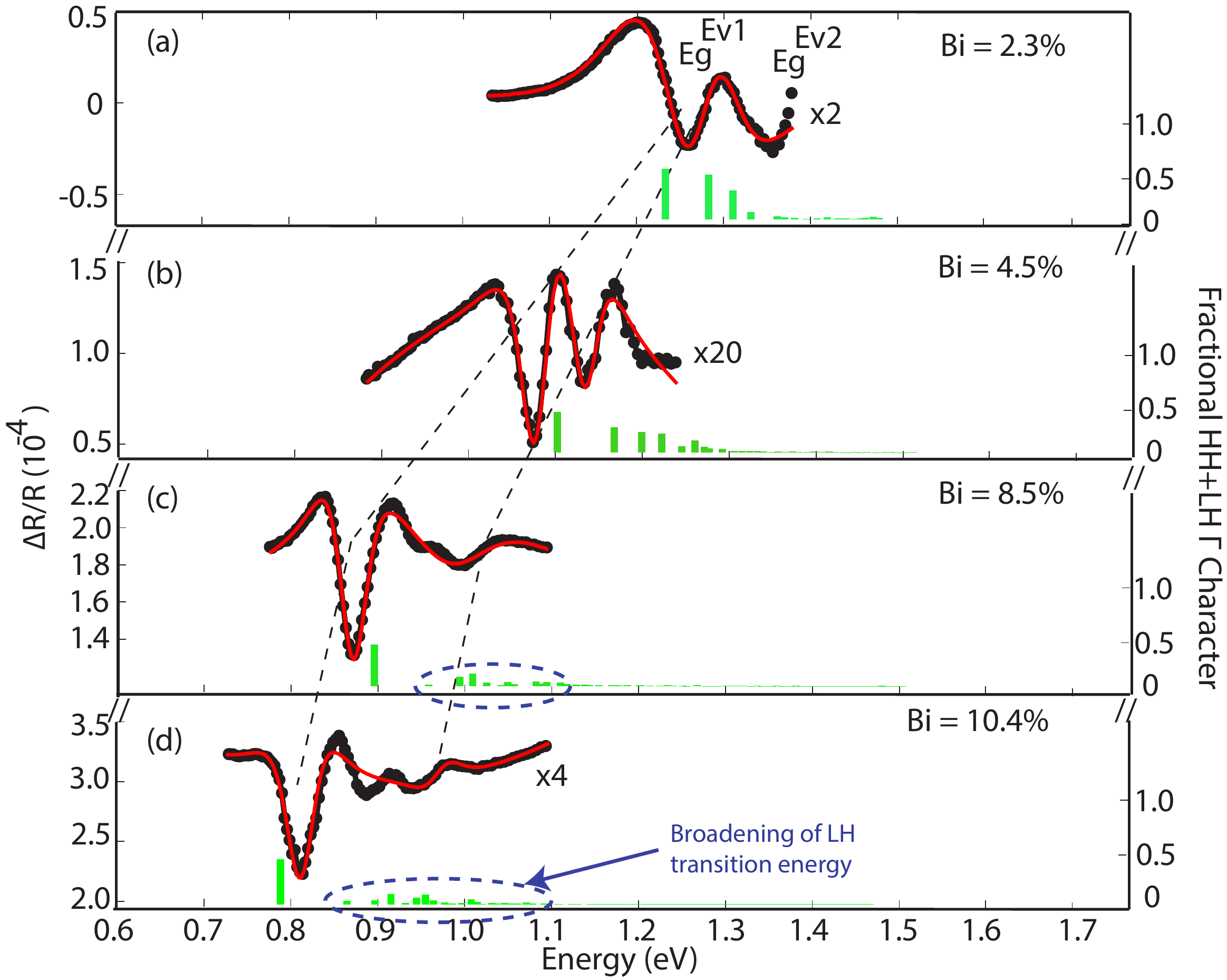}
\caption{Room temperature photo-reflectance (PR) spectra for the four compressively-strained GaBi$_x$As$_{1-x}$ samples, in the region of the fundamental band gap highlighting the E$_g^{E_{v1}}$ = E$_{c1} -$ E$_{v1}$ and E$_g^{E_{v2}}$ = E$_{c1} -$ E$_{v2}$ transitions. The black circles are from measured values and the red lines are the fits obtained by equation \ref{eq:PR_fit}. The green bars are the calculated GaAs HH+LH $\Gamma$-character plots for the alloy valence band states, computed from 8000 atom supercell calculations. The values of the HH and LH related transition energies are selected from the $\Gamma$ character plots as the energies with the highest HH and LH $\Gamma$ character projections, respectively. The dashed lines are a guide to the eye and indicate the PR fitting transition energies.}
\label{fig:HH_LH_PR}
\end{figure}

Fig.~\ref{fig:HH_LH_PR} shows the measured PR spectra (black dots) near to and just above the energy gap for the four GaBi$_x$As$_{1-x}$ samples considered, with compositions $x$ = 2.3\%, 4.5\%, 8.5\%, and 10.4\%. The measured PR data have been fitted using Eq. \ref{eq:PR_fit}, with the fits shown in the figure by the solid red lines. The values of the broadening parameter $\gamma$ used in Eq. \ref{eq:PR_fit} to obtain best fits to the experimental data are listed in table \ref{tab:table1}. These values are directly related both to inhomogeneous broadening of the transition energies due to alloy disorder effects in the grown samples, as well as to any composition fluctuation effects. The fitted $\gamma$ values will be compared to the broadening of our theoretically calculated $\Gamma$-character plots later in this paper. 

For each composition case shown in Fig.~\ref{fig:HH_LH_PR}, cumulative HH+LH $\Gamma$ character spectra are also plotted, using the same energy scale (green bars) and where each peak in the theoretical spectra is associated with a transition between the lowest conduction state and a given valence state. The $\Gamma$-character plots are computed from 8000 atom supercell calculations, with the Bi atoms randomly distributed within each supercell considered. Here we consider only one random distribution of Bi atoms to compare against experiment for each composition. The results here therefore highlight alloy disorder effects such as mixing of alloy valence band states with Bi pair and cluster related states, as a function of the Bi composition. Later in section \ref{sec:random_bi}, we will investigate the second important factor that contributes in the inhomogeneous broadening of the transition energies by performing calculations for four different random distributions of Bi atoms for each grown sample. These later results will show that the specific distribution of Bi atoms within a supercell can further alter the calculated energy spectra by introducing different Bi local environments. This can result in a redistribution of $\Gamma$-character and shifts in the transition energies, thus providing a further contribution to the inhomogeneous broadening of the calculated spectra.

\vspace{2mm}

\begin{figure}
\centering
\includegraphics[width=0.45\textwidth]{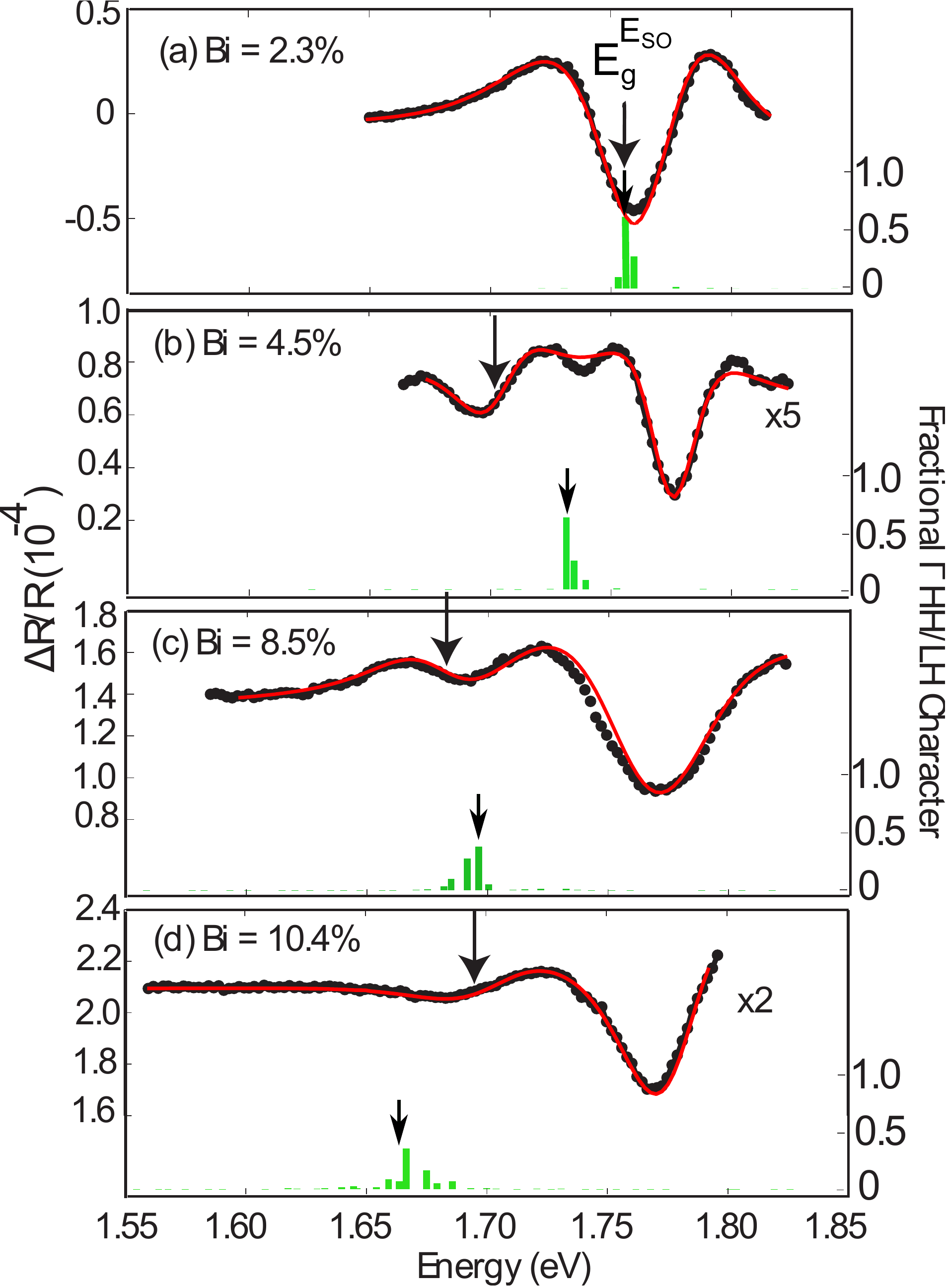}
\caption{Room temperature photo-reflectance (PR) spectra for the four compressively strained GaBi$_x$As$_{1-x}$ samples in the region of the spin-orbit split off transition (E$_g^{E_{SO}}$ = E$_{c1} -$ E$_{SO}$). The black dots are from measured values and the red lines are the fits described by equation \ref{eq:PR_fit}. The green bars are the calculated GaAs SO $\Gamma$-character plots for the alloy valence band states, computed from 8000 atom supercell calculations. The upper arrows in each panel indicate the PR fitting energies and the lower arrows mark the calculated energies from the $\Gamma$ character plots. The prominent feature around 1.77 eV  in the PR spectra is due  to the SO related transition from the GaAs substrate.}
\label{fig:SO_PR}
\end{figure}
\vspace{2mm}
  
We find that the alloy conduction band (CB) $\Gamma$-character is largely concentrated on the lowest energy conduction state of the alloy, which retains more than 95\% $\Gamma$ character for all of the supercells and Bi compositions considered. This is consistent with our previous free-standing GaBi$_x$As$_{1-x}$ supercell calculations, where the conduction band $\Gamma$-character was found to be more than 99.9\% concentrated on the lowest energy state. \cite{Usman_PRB_2011} This confirms that we can use a conventional alloy model to describe the shift in the conduction band edge energy with alloy composition. Therefore, we do not show plots of the CB $\Gamma$-character in this paper, and use the calculated lowest conduction band edge energy, E$_{c1}$, to compute the interband transition energies.  

The two composition dependent features in the PR plots of Fig.~\ref{fig:HH_LH_PR} are assigned for the 2.3\% and 4.5\% samples to the E$_g^{E_{v1}}$ = E$_{c1} -$ E$_{v1}$ and E$_g^{E_{v2}}$ = E$_{c1} -$ E$_{v2}$ transitions. This assignment is supported by the $\Gamma$-character plots, where the lower energy feature is associated with transitions to HH-like valence states ($E_{v1}$) and the higher energy feature with transitions to a state ($E_{v2}$) with significant LH character. The fittings obtained from Eq. \ref{eq:PR_fit} for these samples, shown by the red lines are in good agreement both with the theory and with experiment. 

As the Bi composition is increased to 8.5\% and 10.4\%, increasing disorder, including the presence of a large number of Bi pair and cluster states, results in a large inhomogeneous broadening of the $\Gamma$-character plots over the higher valence band states. This makes the E$_g^{E_{v2}}$ transition difficult to identify in the PR spectra, as can be observed from the poor agreement between the experimental data (black circles) and the fitted values of $\frac{\Delta R}{R}$ (red lines) for these two compositions. The corresponding broadening of the $\Gamma$-character plots is highlighted by the dashed (blue) ovals in Fig.~\ref{fig:HH_LH_PR}, consistent with the difficulty in fitting the high energy feature in the measured PR spectra. It should be noted that the low energy transition involving E$_{c1}$ and E$_{v1}$ does not require an equivalent broadening either in theory or experiment. This is due to the fact that the energy of the E$_{v1}$ state shifts upwards in compressively grown samples, and for high Bi compositions (8.5\% and 10.4\%), it can then still retain significant host matrix HH $\Gamma$ character.  

\begin{figure}
\centering
\includegraphics[width=0.4\textwidth]{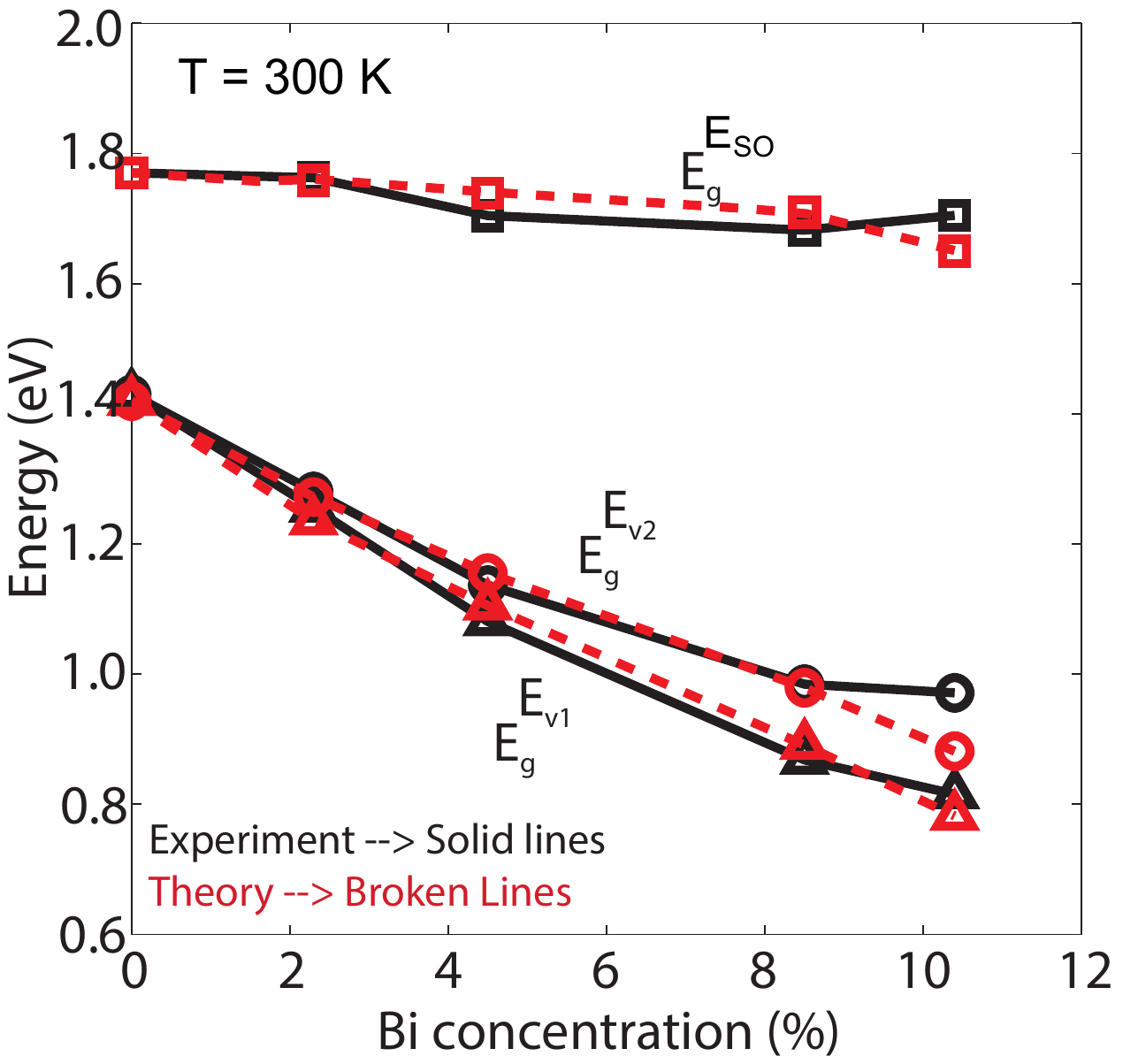}
\caption{Plot of the room temperature transition energies, E$_g^{E_{v1}}$, E$_g^{E_{v2}}$, and E$_g^{E_{SO}}$, obtained from fitting the PR spectra (black solid lines) and calculated from the atomistic simulations (broken red lines). The values are directly extracted from Figs. \ref{fig:HH_LH_PR} and \ref{fig:SO_PR}.}
\label{fig:Band_Edges}
\end{figure}
\vspace{2mm}

Fig.~\ref{fig:SO_PR} shows the measured PR spectra (black circles), fits of Eq. \ref{eq:PR_fit} (red lines), and the calculated GaAs SO band $\Gamma$-character plots (green bars) for the spin-split-off transition energy (E$_g^{E_{SO}}$ = E$_{c1} -$ E$_{so}$). The $\Gamma$-character for the SO band is spread over several states, which are closely spaced in energy. This makes it difficult to choose a particular energy value for the SO transition energy. We select the SO transition energy, illustrated by the lower vertical arrow in each panel, by calculating the weighted average of the energies with their corresponding $\Gamma$-character. The transition energies obtained from fits to the experimental data using Eq. \ref{eq:PR_fit} are shown by the upper vertical arrows in each case. 

Fig.~\ref{fig:Band_Edges} compares the calculated values of the three composition dependent transition energies E$_g^{E_{v1}}$, E$_g^{E_{v2}}$, and E$_g^{E_{SO}}$, with the fitted PR transition energies extracted from Figs.~\ref{fig:HH_LH_PR} and ~\ref{fig:SO_PR}. For the four Bi compositions considered here, our theoretical values of the transition energies are overall in good agreement with the experimental values. The fittings of the E$_g^{E_{v2}}$ and E$_g^{E_{SO}}$ energies for the 10.4\% sample are less good, due to a significant broadening of the LH and SO $\Gamma$-character which makes it difficult to pick a particular energy accurately from the PR data.   

\begin{figure}
\centering
\includegraphics[width=0.4\textwidth]{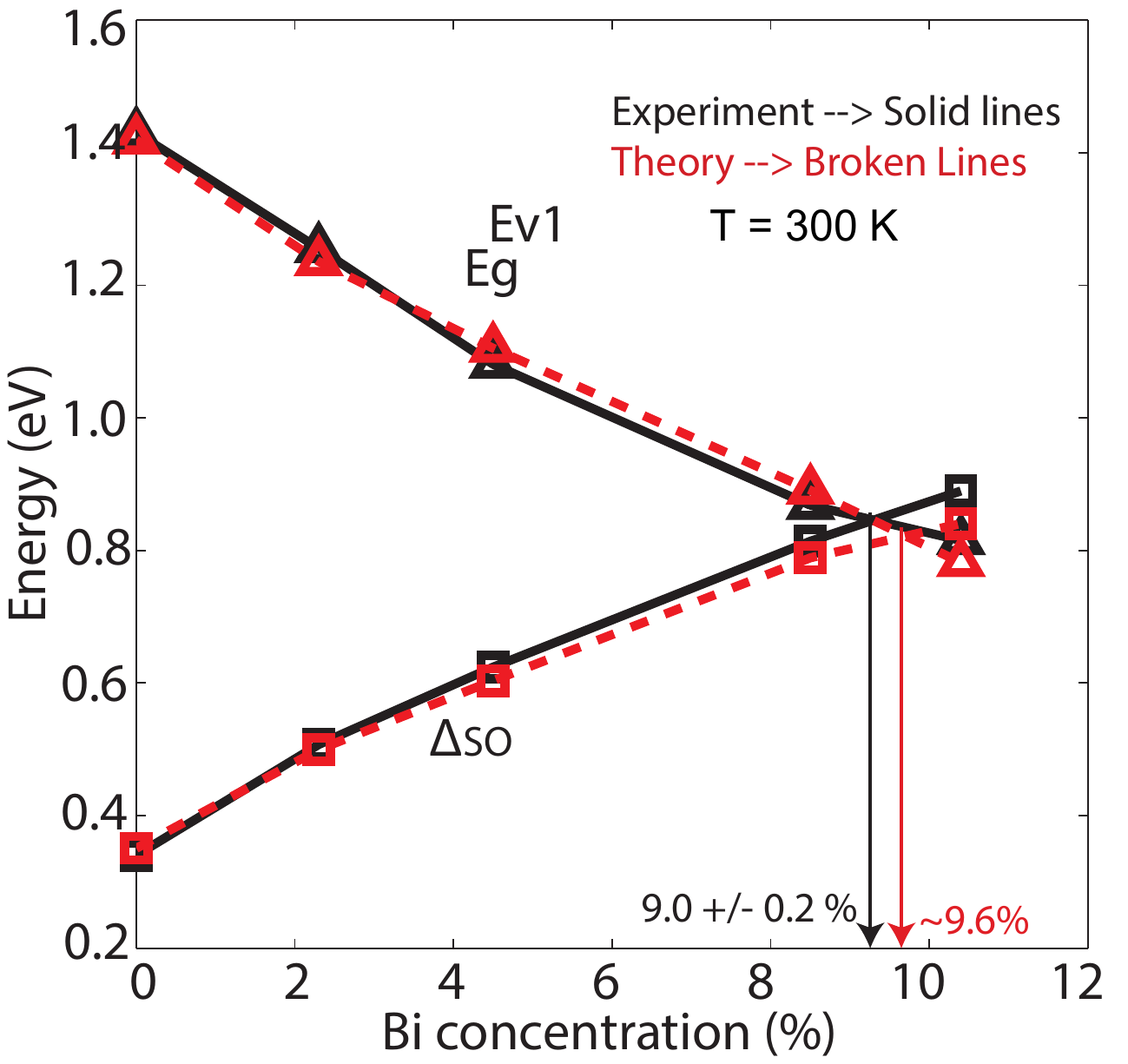}
\caption{Comparison of the experimental and theoretical values of the energy gap  E$_{g}$ and spin-orbit-splitting energy $\Delta_{SO}$ obtained from the results of Fig.~\ref{fig:Band_Edges}. The values of the Bi composition where E$_g$ = $\Delta_{SO}$ obtained from theory and experiment are also shown.}
\label{fig:Band_gap_SO}
\end{figure}

From the results of Fig.~\ref{fig:Band_Edges}, we can further compute the spin-orbit-splitting energy $\Delta_{SO}$ = E$_{v1} -$ E$_{SO}$ as $\Delta_{SO}$ =  E$_g^{E_{SO}} -$ E$_g^{E_{v1}}$. Fig.~\ref{fig:Band_gap_SO} shows the plots of E$_g$ = E$_g^{E_{v1}}$ and of $\Delta_{SO}$ obtained both from our TB calculations and from the experimental PR measurements. The two results are in good agreement and the calculated value of the Bi fraction where E$_g$ = $\Delta_{SO}$  ($\approx$ 9.6\%) is in excellent agreement with the value ($\approx$ 9.0 $\pm$ 0.2\%) obtained from the fits to the PR data. This crossing is technologically important for the design and realisation of photonic devices, due to the possibility of suppressing at higher bismuth concentrations the CHSH Auger recombination loss process, which involves hole excitation from the highest valence band into the spin-split-off band. \cite{Sweeney_ICTON_2011, Sweeney_patent_2010, Broderick_SST_2012}                  

\vspace{2mm}

\subsection{Effect of Supercell Size}
\label{sec:supercell_size}

The size of supercell chosen in any theoretical study of dilute impurity alloys is an important calculational consideration, because it can significantly impact the accuracy of the results. By performing ordered supercell calculations, it has previously been shown that a minimum supercell size of 4096 atoms is required to realize the dilute impurity limit \cite{Usman_PRB_2011, Broderick_PSSb_2012} in GaBi$_x$As$_{1-x}$, GaBi$_x$P$_{1-x}$, and InBi$_x$As$_{1-x}$ alloys. The larger supercell size also increases the accuracy of results by suppressing spurious cell-to-cell interactions between impurities that may arise due to the periodic boundary conditions. \cite{Kent_PRB_2001} 

\begin{figure*}
\centering
\includegraphics[width=0.9\textwidth]{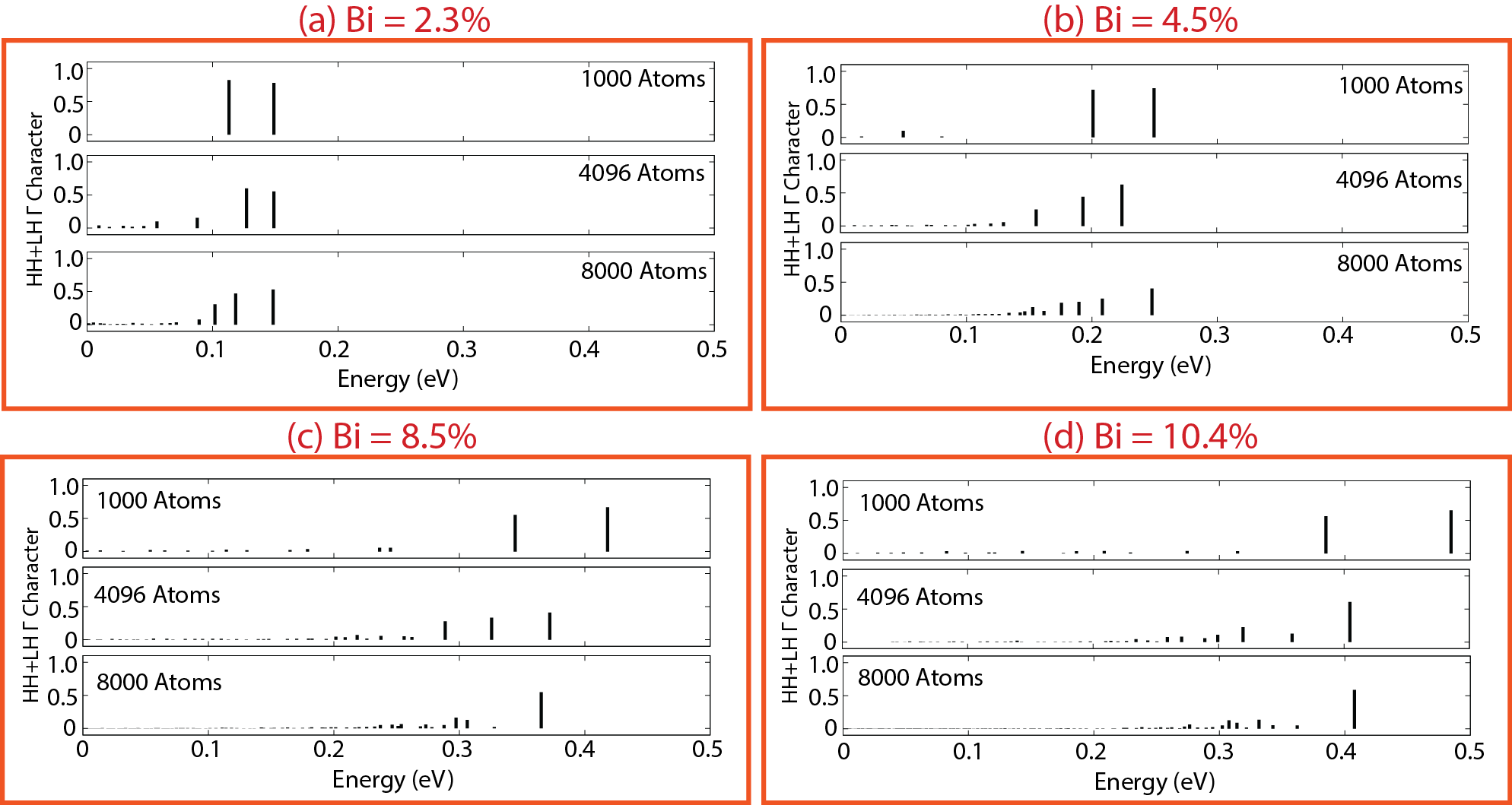}
\caption{For each Bi composition of the experimentally grown samples, we calculate and plot combined HH and LH $\Gamma$ character for three alloy supercells consisting of 1000, 4096, and 8000 atoms. For the same Bi composition, a larger alloy supercell will have more Bi atoms and therefore an increased number of pair and cluster configurations. This is reflected by the wider $\Gamma$ character distribution for larger supercells. The significant overestimation of the alloy VBE energy in 1000 atom supercell calculations for 8.5\% and 10.4\% samples is due to artificially enhanced cell-to-cell interaction of pair and cluster states in the structures considered.}
\label{fig:large_supercell}
\end{figure*}
\vspace{2mm}

In this section, we investigate as a function of the supercell size the impact of alloy disorder on the broadening of $\Gamma$-character plots and on the values of the transition energies. We consider three supercell sizes containing 1000, 4096, and 8000 atoms respectively and compare the HH+LH $\Gamma$-character plots for all four Bi compositions in Fig.~\ref{fig:large_supercell}. 

For the 1000 atom supercells, we calculate two distinct $\Gamma$-character peaks associated with the two highest valence band states for all four Bi compositions considered. This is clearly not consistent with the  measured PR data in Fig.~\ref{fig:HH_LH_PR}, where for the 8.5\% and 10.4\% samples, the higher energy transition corresponding to the E$_{v2}$ state is significantly broadened and becomes hard to define at a single energy value. 

The larger 4096 and 8000 atom supercell calculations are both able to capture this effect where only the highest valence band state E$_{v1}$ is distinct in the $\Gamma$-character plots and the rest of the $\Gamma$-character, in particular the LH-related character, is spread over a large number of valence states, making it difficult to associate a particular energy value with E$_{v2}$. This is because for the 1000 atom supercells, the lower number of Bi atoms results in the formation of fewer pair and cluster states, and the density of states is also lower due to the presence of fewer folded bands near to the GaAs valence band maximum at the $\Gamma$ point. \cite{ Usman_PRB_2011} Therefore, the interband transition involving E$_{v2}$ in theory remains relatively sharp. From these results, we conclude that large supercell calculations are crucial to accurately mimic a dilute impurity alloy.  

Another discrepancy in the 1000 atom calculations is prominent from the $\Gamma$ character plots for the 8.5\% and the 10.4\% compositions in Fig.~\ref{fig:large_supercell}, where the 1000 atom supercell calculations result for both compositions in the E$_{v1}$ state being at significantly higher energy than in the 4096 and 8000 atom calculations. This indicates that the small size of the supercell can overestimate the reduction in the band gap energy at large Bi compositions. Further calculations which we have undertaken demonstrate that there can be a strong variation in the $E_{v1}$ energy obtained from different 1000 atom supercell calculations, where statistical variations in the number of pairs or clusters in the supercell can have a significant impact on the calculated energy spectrum. We will see below that the calculated energy of the highest valence state does not vary significantly between different 8000 atom supercell calculations. Overall, changing the supercell size from 4096 atoms to 8000 atoms provides  a small reduction in the variation between different  $\Gamma$-character plots and therefore a choice can be made between these two supercell sizes, based on the trade-off between desired accuracy and computational time.       

\subsection{Effect of Random Distribution of Bi Atoms}
\label{sec:random_bi}

Another important factor that can significantly influence the theoretically calculated electronic structure of dilute impurity alloys is the particular random distribution of impurity atoms included in the simulated supercells. Different random distributions of Bi atoms will give varying numbers and types of Bi pairs and clusters in the supercell. This can in turn for smaller supercells introduce considerable variations in the $\Gamma$-character plots and in the values of the interband transition energies.  

\begin{figure*}
\centering
\includegraphics[width=0.9\textwidth]{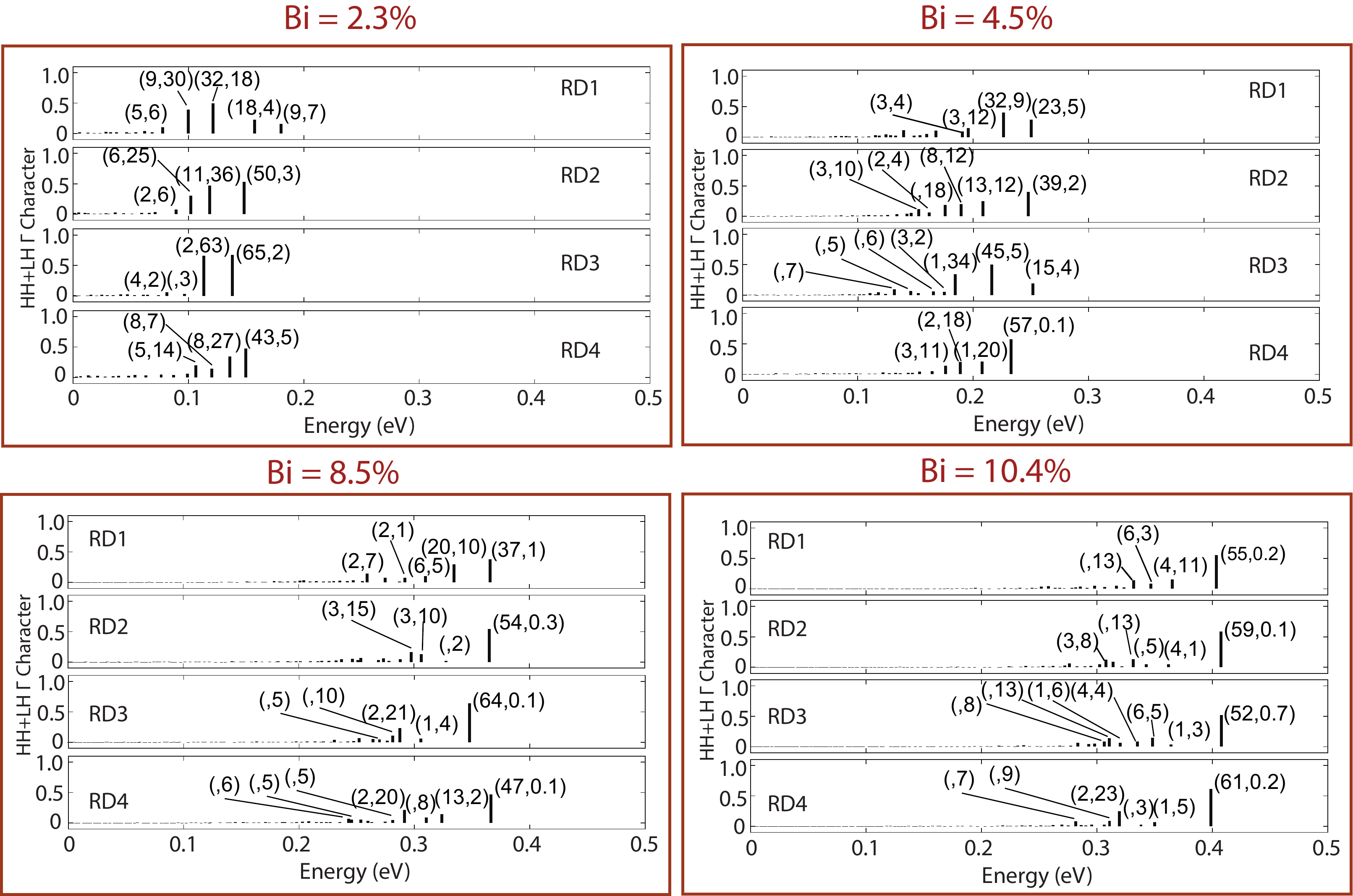}
\caption{For each of the four Bi compositions, we consider four random distributions of Bi atoms in the supercell, labelled as RD1, RD2, RD3, and RD4. Each random distribution introduces different pair and cluster states and therefore exhibits different broadening of the combined HH and LH $\Gamma$-character. The calculated data are shown for 8000 atom supercell calculations. We also specify the values of individual HH and LH contributions in the few highest energy bars of the plot as (HH,LH). If (,LH) or (HH,) is mentioned for a particular bar, that indicates that HH or LH contribution in that bar is less than 0.1\%. The zero of energy is taken in all cases at the GaAs valence band maximum.}
\label{fig:Atomic_randomness}
\end{figure*}
\vspace{2mm} 

In this section, we perform a series of 8000 atom supercell calculations for each of the four Bi compositions to investigate the effect of statistical variations on the electronic properties of the alloy. For each composition, we consider four different random distributions (RDs) of Bi atoms labelled as RD1, RD2, RD3, and RD4, which are obtained by varying the seed values of a random number generator that determines the nature of anion (either Bi or As) at a given atomic location in the supercell. By using different random number seed values, we ensure that all of the four supercells (RD1-RD4) have different arrangements of the Bi atoms, resulting in different numbers and types of Bi pairs and clusters.

Fig.~\ref{fig:Atomic_randomness} plots the sum of the HH and LH $\Gamma$-character, $G_{\Gamma}(E)$, for the four random distributions of Bi atoms considered at each composition. We specify the separate contributions from the GaAs HH and LH states in the label (HH,LH) for the highest few states, in order to highlight the distribution of the HH and LH character over the alloy states. It should be noted that (,LH) and (HH,) refers to cases where a particular alloy state has less than 0.1\% HH or LH character, respectively.  

The impact of variations between different random distributions is more prominent for the two lower Bi compositions, 2.3\% and 4.5\%. This can be mainly attributed to two reasons: (i) For the lower Bi compositions, there is a significantly smaller number of pairs and clusters in a supercell, and therefore the impact of any variation in their type or number is more pronounced. Similar conclusions were presented earlier \cite{Usman_PRB_2011} based on free-standing supercell calculations, where the addition of a single pair or a single Ga-centred 3 Bi cluster in a 4096 atom Ga$_{2048}$Bi$_M$As$_{2048-M}$ supercell ($M$=2, 3) resulted in very large variations in the valence band edge energies. (ii) The highest valence band edge energy is very close to the energies of the pair/cluster states and the stronger state mixing which this allows therefore results in the broadened $\Gamma$-character distributions. This behaviour is confirmed by close examination of RD1 for 2.3\% Bi and RD1 and RD3 for 4.6\% Bi. These are the distributions which give rise to the highest valence band maximum energies, and also have the lowest $\Gamma$ character associated with the highest state, reflecting the contribution of (higher energy) pair or cluster states in each of these cases. The large $\gamma_{HH}$ value required to fit the PR spectrum of the 2.3\% sample may then include a contribution due to mixing of the host $\Gamma$ states with pair and cluster states which lie close to the band edge at this composition.   

For the larger Bi compositions, 8.5\% and 10.4\%, the highest valence band edge energy increases well above the energies of the pair/cluster states and therefore a sharp peak is observed for $E_{v1}$ in the $\Gamma$-character plots. This is quite evident for the 10.4\% composition case, where 55\%, 59\%, 52\%, and 61\% HH character is found on the highest valence band state for the four random distributions. For these compositions, the LH $\Gamma$-character is in each case distributed over a broad range of valence band states, so that it then proves very difficult in the 10.4\% cases to identify any single state as the LH-like valence band maximum, consistent with the difficulty in fitting the PR spectra in this energy range.

\subsection{Summary of the Alloy Disorder Effects and Comparison with the PR fits}
\label{sec:summary_fits}

In the previous two subsections \ref{sec:supercell_size} and \ref{sec:random_bi}, we have investigated the calculated impact of alloy disorder when varying two important parameters: supercell size and the assumed (random) distribution of impurity atoms. Both of these factors have been found to significantly impact the calculated electronic structure of GaBi$_{x}$As$_{1-x}$ alloys, and therefore require careful attention. Below, we summarize our results and the key physical insights gained in the previous subsections concerning the electronic properties of the alloy:

\begin{itemize}

\item A very small supercell size can result in  artificially enhanced alloy disorder effects on the band edge energies, and can therefore introduce significant errors in the calculated band gap energy. Small supercells also fail to correctly capture the broadening of LH character observed in experimental PR data for higher Bi compositions. Our calculations show that a supercell size of 4096 or larger is required to accurately model dilute bismide alloys.

\item The effect of atomistic randomness remains important at low Bi compositions (2.3\% and 4.5\%) even in larger supercells, where variations in the local Bi environments can significantly impact the calculated distribution of $\Gamma$-character, consistent with the large $\gamma_{HH}$ value required to fit the 2.3\% PR spectrum.

\item At higher Bi compositions (8.5\% and 10.4\%), the effect of atomistic randomness is less marked at the valence band maximum, because the E$_{v1}$ energy is shifted well above the pair/cluster state energies. 

\item Overall, the broadening of the theoretically calculated $\Gamma$-character plots correlates well with the values of the broadening parameter $\gamma$ (provided in table \ref{tab:table1}) used to fit the PR spectra via Eq. \ref{eq:PR_fit}. The large values of the $\gamma$ parameter used to fit the PR data for the 2.3\% sample may include a contribution due to composition variations in the grown sample, as well as a contribution due to state mixing with pair and cluster states. For the three higher composition samples,  a constant value of $\gamma_{HH} $ can be chosen to fit the HH-related transition. This is consistent with the theoretically computed $\Gamma$-character plots, where the HH-character is largely present on the alloy VBE state and does not show much spreading even for different random distributions of the Bi atoms, as evident from the plots in Fig. \ref{fig:Atomic_randomness}. 

\item The value of the $\gamma_{SO}$ broadening parameter in table \ref{tab:table1} for the SO band slightly increases with Bi composition. This is consistent with the $\Gamma$-character plots shown in Fig. \ref{fig:SO_PR}, which show an increase in broadening of the $\Gamma$-character with increasing composition. In order to quantify this effect, we estimate the calculated broadening of the SO $\Gamma$ character by calculating the energy range $\bigtriangleup E$ around E$_{SO}$ which includes 70\% of the SO character. The values obtained for 2.3\%, 4.5\%, 8.5\%, and 10.4\% Bi compositions are $<$1 meV, 2.7 meV, 5.7 meV, and 13.8 meV, clearly indicating an increasing broadening of SO $\Gamma$-character with increasing Bi composition, consistent with conventional alloy disorder effects. 

\item The behaviour of the $\gamma_{LH}$ parameter values for the LH-related transition energy is also well explained by the theoretical analysis. Table ~\ref{tab:table1} shows a significant increase in $\gamma_{LH}$ for the 8.6\% Bi sample, while a good fit could not be obtained to the experimental data for the 10.4\% sample, due to the presence of extra features in the measured data. This is consistent with our TB calculations that show a significant broadening of the LH $\Gamma$-character over a series of alloy states at higher compositions, making it very difficult to reliably choose a particular energy value for the E$_{v2}$-related transition energy in the 10.4\% sample.

\end{itemize}

\subsection{Conduction Band Edge Deformation Potential and Type-I Band Edge Alignments}
\label{sec:band_alignment}

\begin{figure}
\centering
\includegraphics[width=0.44\textwidth]{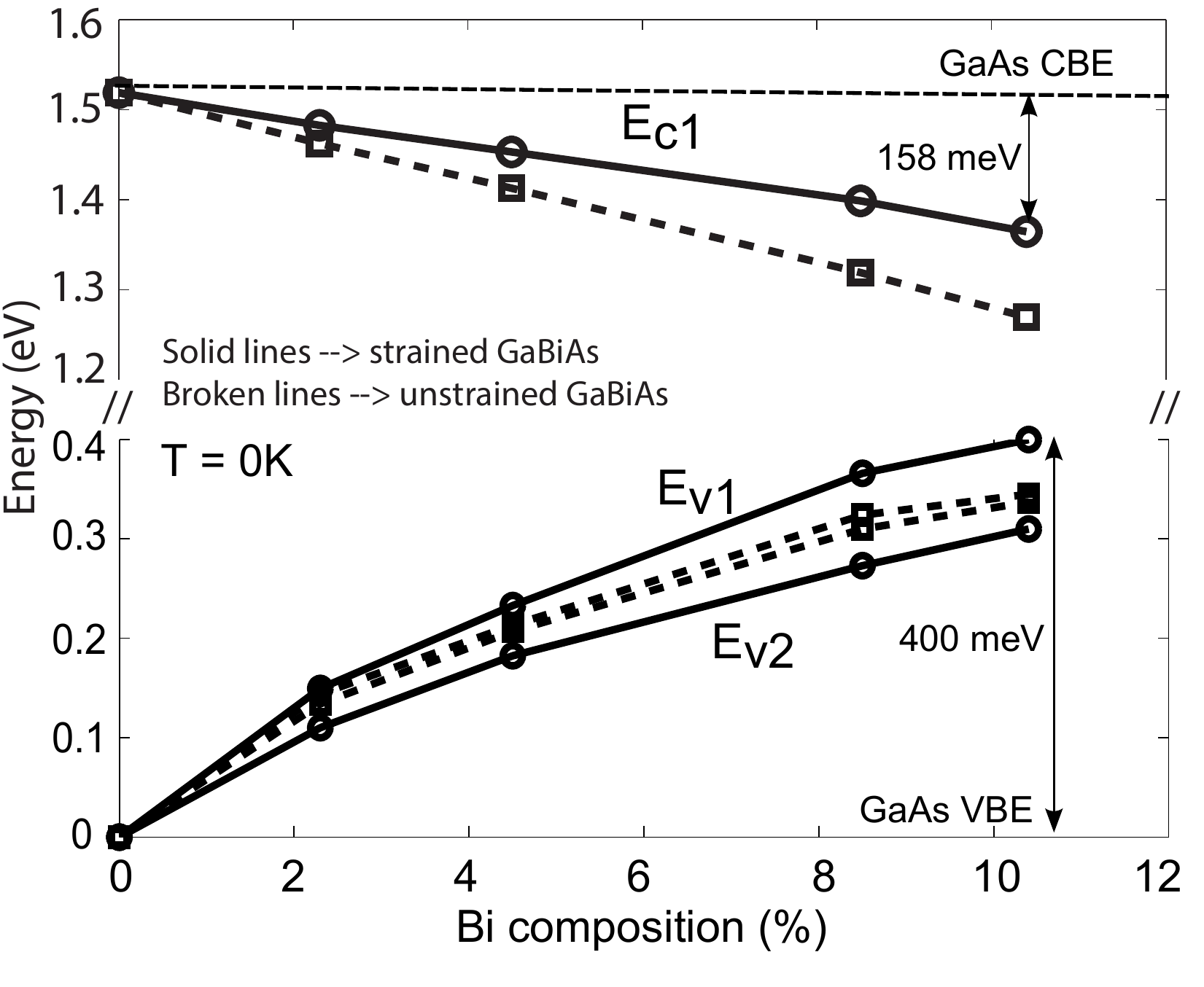}
\caption{The Bi composition dependent shifts in the band edges energies E$_{c1}$, E$_{v1}$, and E$_{v2}$ of the GaBi$_x$As$_{1-x}$ alloy are shown, obtained from the compressively strained and free-standing 8000 atom supercell calculations at low temperature. The strain shifts the lowest conduction band edge to higher energy, reducing its slope from 28 meV/\%Bi to 15 meV/\%Bi.  The splitting between the two highest valence band edges is also increased due to compressive biaxial strain.}
\label{fig:Band_alignment}
\end{figure}

Fig.~\ref{fig:Band_alignment} compares the Bi composition dependent band edge energies for the lowest conduction band edge (E$_{c1}$) and the highest two valence band edges (E$_{v1}$ and E$_{v2}$) for the compressively strained and unstrained (free-standing) GaBi$_x$As$_{1-x}$ supercells, calculated from our atomistic tight binding model. The compressive hydrostatic component of strain pushes the conduction band edge energy upwards and the biaxial strain leads to an increasing splitting between the two highest valence band states with increasing $x$. 

We calculate a strain induced net decrease in E$_{c1}$ of 13 meV per \% Bi. Using $\delta E_{c} = a_{c} \epsilon_{H}$ where $\delta E_{c}$ is the change in the conduction band edge energy and $\epsilon_{H}$ is the hydrostatic strain, we can estimate the value of CB deformation potential $a_{c}$ for the GaBi$_{x}$As$_{1-x}$ alloy. Using the values from Fig. \ref{fig:Band_alignment} at 10.4\% Bi composition, $\delta E_{c}$ = 0.12 eV. The hydrostatic strain is $\epsilon_{H}$ $\approx$ -0.013. Therefore the calculated value of $a_{c}$ for the compressively strained GaBi$_{x}$As$_{1-x}$ is -9.4 eV, which is comparable to, but of larger magnitude than the values -7.17 eV and -5.08 eV reported earlier for GaAs and InAs materials. \cite{Walle_PRB_1989}           

The determination of the alignment of conduction and valence bands is critical for the design and optimization of optoelectronic devices. The band alignment at the GaBi$_x$As$_{1-x}$/GaAs interface remains uncertain in the literature with reports of type I, \cite{Usman_PRB_2011, Nadir_APL_2012, Pettinari_APL_2008} type II, \cite{Alberi_PRB_2007} and nearly flat \cite{Tominaga_ISLC_2010} conduction band offsets. We calculate here that  the lowest conduction band edge energy moves up by $\approx$ 13 meV per \% Bi with  increasing Bi composition. This upward shift is not large enough to overcome the $\approx$ 28 meV per \% Bi decrease in energy due to the conventional alloy-type shift. Therefore for all the bismuth compositions considered, the conduction band edge of GaBi$_x$As$_{1-x}$ remains  below the conduction band edge of the GaAs barrier material. The valence band edge, on the other hand, increases in energy due to the BAC interaction effect in the free-standing supercells and compressive strain further enhances this upward shift, so the top most valence band edge of the GaBi$_x$As$_{1-x}$ alloy is always well above the valence band edge in the GaAs barrier material. At 10.4\% Bi composition for which $\bigtriangleup_{SO} > E_{g}$, we calculate band offsets of 160 meV and 400 meV for the conduction and valence band edges with respect to the GaAs bulk band edges, respectively. From this discussion, we conclude that GaBi$_x$As$_{1-x}$ alloys grown pseudomorphically on a GaAs substrate have a type I band alignment at the GaAs barrier interface, in accordance with more recent experimental findings. \cite{Batool_JAP_2012, Nadir_APL_2012} The increase in the calculated value of $a_c$ for GaBi$_x$As$_{1-x}$ and the calculated type I band alignment are, in particular, consistent with the results of hydrostatic pressure measurements undertaken on GaBi$_x$As$_{1-x}$/GaAs LED structures \cite{Nadir_APL_2012} These measurements found that the electro-luminescence intensity from GaBiAs decreased with increasing pressure, accompanied by an increase in the luminescence from the GaAs layer, consistent with increased carrier overflow with increasing pressure, owing to a type I offset and a larger conduction band deformation potential in the GaBiAs layer, than in the surrounding GaAs layers.

\vspace{2mm}


\section{Conclusions}
\label{sec:Conclusions}

In summary, we have investigated alloy disorder effects on the electronic structure of compressively strained GaBi$_x$As$_{1-x}$ samples grown on a GaAs substrate by comparing our atomistic tight binding results with room temperature photo-modulated reflectance measurements. Our theoretical results are in good agreement with the experimental data and reproduce the crossover at 9\% Bi composition, beyond which $\Delta_{SO}$ exceeds the energy gap E$_g$. Our theoretical results explain that for the higher Bi composition samples with Bi fractions of 8.5\% and 10.4\%, the presence of disorder, including a large number of Bi pair and cluster states close to the host matrix light-hole band edge, signifcantly increases the mixing of the LH band edge with Bi-related resonant states, and therefore makes it difficult to identify a well defined LH band edge energy at higher compositions in the strained alloy. This is consistent with the poor fit obtained to the PR spectra above the band edge at these compositions. We have also analysed the impact of supercell size and of statistical variations in the  Bi distribution on the calculated electronic structure, showing that large ($\ge 4000$ atom) supercells are required to describe the alloy band structure, and to understand the experimentally observed inhomogeneous broadening of the transition energies. By comparing the band edge shifts in compressively strained supercells with those calculated for free-standing (unstrained) supercells, we have derived composition dependent deformation potentials for the GaBi$_x$As$_{1-x}$ alloy band edges. Finally, our calculations suggest a type I band alignment at the interface between GaBi$_x$As$_{1-x}$ and GaAs, in accordance with recent experimental analysis. Overall the calculations are in good agreement with the PR measurements and provide useful new insight for the development of this interesting new class of semiconductor materials.


\section*{Acknowledgements}
This work is supported by the European Union Seventh Framework Programme (BIANCHO; FP7-257974), the Irish Research Council (RS/2010/2766), and the UK Engineering and Physical Sciences Research Council (EP/H005587/1, EP/G064725/1). T.J.C.H. thanks U.T.M. and M.O.H.E. for grants 01H55 and 4D301. ZB acknowledges the Islamia University of Bahawalpur, Pakistan for an FDP studentship and partial support from the Kwan trust. MU acknowledges the use of computational resources from the National Science Foundation (NSF) funded Network for Computational Nanotechnology (NCN) through http://nanohub.org. NEMO 3D based open source tools are available at: \url{https://nanohub.org/groups/nemo_3d_distribution}.


\bibliographystyle{apsrev4-1}
\providecommand{\newblock}{}

\end{document}